\documentclass[twocolumn,prl,letter,showpacs,preprintnumbers,amsmath,amssymb,superscriptaddress]{revtex4}

\usepackage{graphicx}
\usepackage{graphicx,epsf} 
\usepackage{dcolumn}
\usepackage{bm}
\usepackage{latexsym}

\newcommand{\kav}{\langle k \rangle}

\begin{document}

\title{Optimization of Network Robustness to Waves of Targeted and Random Attacks}

\author{T.~Tanizawa}
\email{toshi@argento.bu.edu}
\affiliation{Center for Polymer Studies and Dept.\ of Physics, Boston University, Boston, MA 02215, USA}
\affiliation{Department of Electrical Engineering, Kochi National College of Technology,
Monobe-Otsu 200-1, Nankoku, Kochi 783-8508, JAPAN}
\author{G.~Paul}
\affiliation{Center for Polymer Studies and Dept.\ of Physics, Boston University, Boston, MA 02215, USA}
\author{R.~Cohen}
\affiliation{Minerva Center and Department of Physics, Bar Ilan University Ramat Gan 52900, Israel}
\author{S.~Havlin}
\affiliation{Center for Polymer Studies and Dept.\ of Physics, Boston University, Boston, MA 02215, USA}
\affiliation{Minerva Center and Department of Physics, Bar Ilan University Ramat Gan 52900, Israel}
\author{H.~E.~Stanley}
\affiliation{Center for Polymer Studies and Dept.\ of Physics, Boston University, Boston, MA 02215, USA}
 
\begin{abstract}

We study the robustness of complex networks to multiple waves of simultaneous (i)
targeted attacks in which the highest degree nodes are removed and (ii)
random attacks (or failures) in which fractions $p_t$ and $p_r$
respectively of the nodes are removed until the network collapses.  We
find that the network design which optimizes network robustness has a
bimodal degree distribution, with a fraction $r$ of the nodes having
degree $k_2= (\kav - 1 +r)/r$ and the remainder of the nodes having
degree $k_1=1$, where $\kav$ is the average degree of all the nodes.  We
find that the optimal value of $r$ is of the order of $p_t/p_r$ for $p_t/p_r\ll 1$.

\end{abstract}


\maketitle

Recently, there has been much interest in the resilience of real-world
networks to random attacks or to attacks targeted on the highest degree nodes
\cite{Albert,Paxon,Cohen2000,Callaway,Cohen2001,Cohen2002,Valente2004,Paul2004}.
Many real-world networks are robust to random attacks but vulnerable to
targeted attacks.  It is important to understand how to design networks
which are optimally robust against both types of attacks, with
examples being terrorist attacks on physical networks and attacks by
hackers on computer networks.  Studies to date \cite{Valente2004,Paul2004}
have considered only the case in which there was only one type of attack on a given
network --- that is, the network was subject to either a random attack or
to a targeted attack but not subject to different types of attack simultaneously.

A more realistic scenario is one in which a network is subjected to
simultaneous targeted and random attacks.  This scenario can be modeled
as a sequence of ``waves'' of targeted and random attacks which remove
fractions $p_t$ and $p_r$ of the original nodes, respectively. The ratio
$p_t/p_r$ is kept constant while the individual fractions $p_t$ and
$p_r$ approach zero.  After some time (after $m$ waves of random and
targeted attacks) the network will become disconnected; at this point a
fraction $f_c = m(p_t + p_r)$ of the nodes have been removed.
This $f_c$ characterizes the network robustness.
The larger $f_c$, the more robust the network is. 
We propose in this Letter a mathematical approach to study such simultaneous
attacks and find the optimal network design one which maximizes $f_c$
In our optimization analysis, we compare the robustness of 
networks which have the same ``cost'' of construction and
maintenance, where we define cost to be proportional to the average
degree $\langle k \rangle$ of all the nodes in the network.

We study mainly two types of random networks:

\begin{itemize}

\item[{(i)}] {\it Scale-free networks}.  Many real world computer,
  social, biological and other types of networks have been found to be
  scale free, i.e., they exhibit degree distributions of
  the form $P(k) \sim k^{-\lambda}$ %
  ~\cite{Bar99,Faloutsos,Barabasi,Broder,Ebel,Redner,Jeong,Mendes,RPT}.
  For large scale-free networks with exponent $\lambda$ less than 3, for
  random attacks essentially all nodes must be removed for the network
  to become disconnected~\cite{Cohen2000,Callaway}. On the other hand,
  because the scale free distribution has a long power-law tail
  (i.e. hubs with large degree), the scale-free networks are very vulnerable
  with respect to targeted attack.

\item[{(ii)}] {\it Networks with bimodal degree distributions.}  For
  resilience to single random or single targeted attacks, certain
  bimodal distributions are superior to any other network~\cite{Valente2004,Paul2004}.
  Here we ask if these networks are also most resilient
  to multiple waves of both random and targeted attacks.

\end{itemize}

We present the following argument that suggests that the degree
distribution which optimizes $f_c$ is a bimodal distribution in which a
fraction $r$ of the nodes has degree
\begin{equation}
k_2= {\kav - 1 +r \over{r}} 
\label{eq0}
\end{equation}
and the remainder has degree $k_1=1$ and we show that $r$ is of the
order of $p_t/p_r$.  To optimize against random removal, we maximize the
quantity $\kappa \equiv \langle k^2 \rangle / \langle k \rangle$, since
for random removal the threshold is ~\cite{Cohen2000}
\begin{equation}
f_c^{\textrm{rand}}=1-{1\over{\kappa -1}}.
\label{eq6}
\end{equation}
Since we keep $\kav$ fixed, $\kappa$ is just the variance of the degree
distribution and is maximized for a bimodal distribution in which the
lower degree nodes have the smallest possible degree $k_1=1$ and the
higher degree nodes have the highest possible degree consistent with
keeping $\kav$ fixed, $k_2= (\kav - 1 +r)/r$.  Thus, $k_2$ is maximized
when $r$ assumes its smallest possible value, $r=1/N$.  On the other
hand, if all of the high degree nodes are removed by targeted attacks,
the network will be very vulnerable to random attack.  So we want to
delay as long as possible the situation in which all of the high degree
nodes are removed by targeted attacks---which argues for not choosing
$r$ as small as possible but choosing $r$ such that some high
connectivity nodes remain as long as there are some low connectivity
nodes.  Such a condition is achieved when $r$ is of the order of
$p_t$/$p_r$.

\begin{figure}[thb]
\includegraphics[width=7.5cm]{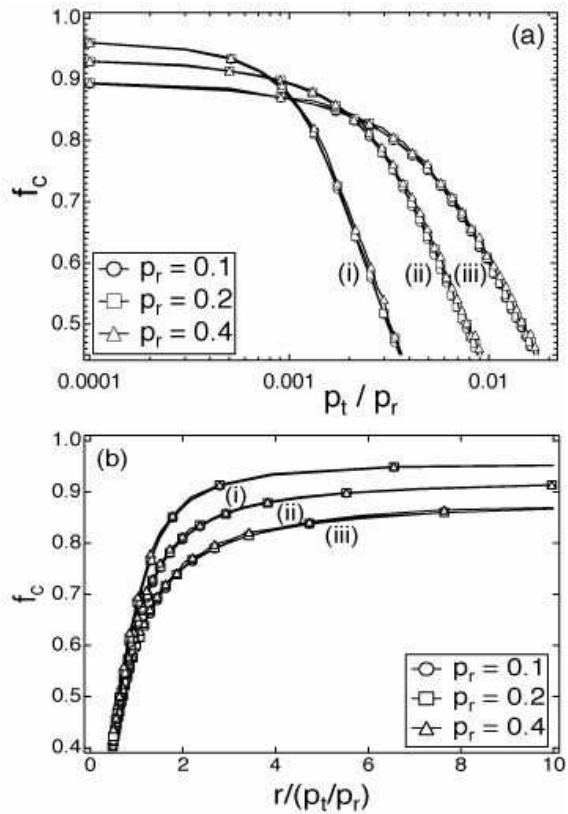}
\caption{(a) The threshold $f_c$ of three bimodal networks with $\langle
   k \rangle = 3$, with (i) $r = 2 \times 10^{-3}$ and $k_2 = 200$, (ii)
   $r = 5 \times 10^{-3}$ and $k_2 = 90$, and (iii) $r = 10^{-2}$ and
   $k_2 = 50$.  The results are plotted as a function of the ratio $p_t
   / p_r$ for three fixed values of $p_r$.  These plots show that the
   values of the threshold are dependent only on the ratio $p_t/p_r$ and
   independent of the value of $p_r$ itself. (b) Scaled plot of the data
   in (a).  The data show that the plots collapse in the region where
   $r/\left( p_t/p_r \right) \lesssim 1$.}
\label{fig:prIndependent}
\end{figure}

The method we employ for determining the threshold makes use of the following:
the general condition for a random network to be globally
connected is~\cite{Cohen2000, Cohen2001, Cohen2002}
\begin{equation}
\kappa = \frac{\langle k^2 \rangle}{\kav} \ge 2.
\end{equation}
Random removal of a fraction $p_r$ of nodes from a network with
degree distribution $P_0(k)$ results in a new degree distribution \cite{Cohen2000}
\begin{equation}
P(k) = \sum_{k_0 = k}^K P_0(k) \binom{k_0}{k} \left( 1 - p_r \right)^k
p_r^{k_0 - k},
\label{eq:random attacks}
\end{equation}
where $K$ is the upper cutoff of the degree distribution.
Targeted removal of a fraction $p_t$ of the highest degree nodes
reduces the value of upper cutoff $K$ to $\tilde{K}$,
which is implicitly determined by the equation
\begin{equation}
p_t = \sum_{k = \tilde{K}}^K P_0(k).
\label{eq:Ktilde}
\end{equation}
The removal of high degree nodes causes another effect.
Since the links that lead to removed nodes are eliminated,
the degree distribution also changes.
This effect is equivalent to the random removal of a fraction of $\tilde{p}$ nodes where
\begin{equation}
\tilde{p} = \frac{\sum_{k = \tilde{K}}^K k P_0(k)}{\langle k \rangle_0}.
\label{eq:ptilde}
\end{equation}
The average $\langle k \rangle_0$ is taken over the degree distribution
before the removal of nodes~\cite{Cohen2001}.
Equation~(\ref{eq:random attacks}) with $p_r$ replaced by $\tilde{p}$ can then
be used to calculate the effect of the link removal.
Starting with a certain initial degree distribution,
we recursively calculate $P(k)$ alternating between random and targeted attack
using Eqs.~(\ref{eq:random attacks}), (\ref{eq:Ktilde}), and (\ref{eq:ptilde}),
and calculate $\kappa$ after each wave of attacks.
When $\kappa < 2$ global connectivity is lost and $f_c = m (p_r + p_t)$
where $m$ is the number of waves of attacks performed.

We begin our study by first establishing numerically that, for small values of
$p_t$, $p_r$ and $p_t/p_r$, the threshold $f_c$ depends only on $p_t/p_r$.  In
Fig.~\ref{fig:prIndependent}(a), we plot the threshold $f_c$ of a
network with a bimodal degree distribution with $\langle k \rangle = 3$
for various values of $p_r$ and $r$ as a function of the ratio $p_t /
p_r$.  The collapse of the plots with the same $r$ but different $p_r$
shows that the values of threshold are essentially independent of the
value of $p_r$ itself but depend only on the ratio $p_t/p_r$~\cite{noteX}.

In Fig.~\ref{fig:prIndependent}(b) we plot $f_c$ against the scaled
variable $r/(p_t/p_r)$. We see that the plots for different values of
$r$ collapse, indicating that only the scaled variable $r/(p_t/p_r)$ is
relevant~\cite{noteA}.

\begin{figure}[thb]
\includegraphics[width=7.5cm]{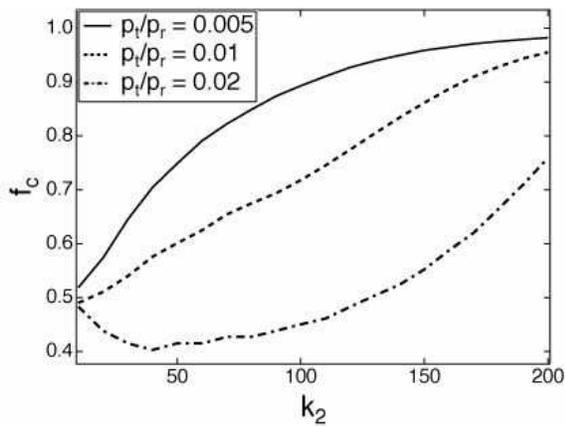}
\caption{The threshold $f_c$ versus $k_2$ for a bimodal network with
   $\langle k \rangle = 3$ and $r = 10^{-2}$ for three values of $p_t /
   p_r$.  The value of $p_r$ is fixed at 0.02.  For each value of
   $p_t/p_r$, the thresholds take their maximum values at the maximum
   $k_2$ (obtained when $k_1=1$).}
\label{fig:vark2}
\end{figure}

Next we study the dependence of $f_c$ on $k_2$. As seen in Fig.~\ref{fig:vark2},
as expected the maximum values of $f_c$ for various
values of $p_t/p_r$ are obtained when $k_2$ is maximum (i.e., when
$k_1=1$, see Eq.~(\ref{eq0}).).

\begin{figure}[thb]
\includegraphics[width=7.5cm]{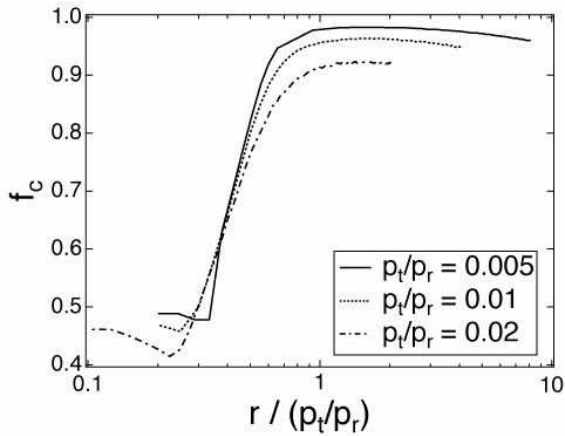}
\caption{The threshold $f_c$ versus the scaled parameter $r/ (p_t /
   p_r)$ for a bimodal network with $\kav=3$ and $k_2$ maximum (i.e.,
   $k_1=1$).}
\label{fig:kAv3Scaled}
\end{figure}

\begin{figure}[thb]
\includegraphics[width=7.5cm]{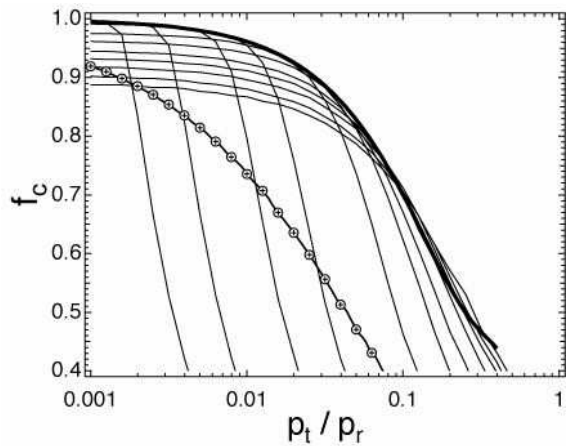}
\caption{The threshold $f_c$ versus $p_t/p_r$.  The topmost (thickest)
   curve is for a bimodal network with $\kav=3$ with $k_1=1$ and with
   $r$ optimized by Eq.~(\ref{eq:ropt}) for each value of $p_t/p_r$. The
   values of the threshold for the same bimodal network with $k_1=1$
   when we fix $r$ independent of $p_t/p_r$ are plotted in thin curves.
   The values of $r$ are $r = 0.001, 0.002, 0.005, 0.01, 0.03, 0.05,
   0.07, 0.09, 0.11, 0.13$, and $0.15$, from left to right.  The curve
   marked with crossed circles ($\oplus$) is a plot of the threshold
   values for a scale-free network with $\kav=3$, $N=10^4$, and with
   exponent chosen for each $p_t/p_r$ to optimize the threshold. Note
   that the thresholds for bimodal networks with $0.03 \lesssim r
   \lesssim 0.09$ are always more robust than the optimized scale-free
   network.}
\label{final2delta}
\end{figure}

We are now in a position to determine the value of $r$ which optimizes
$f_c$, $r_{\textrm{opt}}$.
In Fig.~\ref{fig:kAv3Scaled}, we plot $f_c$ as a function of the
scaled parameter $r/ (p_t / p_r)$ with $k_2$ set to the maximum value
possible for each value of $r$. We note that there is a transition at a
well-defined value of $r/(p_t/p_r)$ at which $f_c$ increases rapidly to
a shallow maximum $f_c^{\textrm{opt}}$ at $r_{\textrm{opt}}/(p_t/p_r) \approx 1.7$.
This value of $r_{\textrm{opt}}/(p_t/p_r)$ is valid for $p_t/p_r \ll 1$.
In order to determine $r_{\textrm{opt}}/(p_t/p_r)$ over a wider range,
we make extensive numerical calculations for $10^{-3} < p_t/p_r < 0.1$.
For each value of $p_t/p_r$, we calculate the value $r_{\textrm{opt}}/(p_t/p_r)$
where $f_c$ takes its maximum value and find
\begin{equation}
\frac{r_{\textrm{opt}}}{p_t/p_r} \approx
1.7 - 5.6 \left( \frac{p_t}{p_r} \right)
+ O\left( \frac{p_t}{p_r} \right)^2
\label{eq:ropt}
\end{equation}
within the range of our calculation.  For larger values of $p_t/p_r$,
$r_{\textrm{opt}} = 1$ and from Eq.~(\ref{eq0}) all nodes have degree $\kav$.
In Fig.~\ref{final2delta}, we plot the values of the optimal threshold
$f_c^{\textrm{opt}}$ by a thick solid curve. 

In Fig.~\ref{final2delta} we also plot the values of the threshold $f_c$
for the same bimodal network but we fix $r$ independent of $p_t/p_r$.
We see that these configurations are not significantly less robust than
the optimal configuration. Thus, even if we do not know the ratio $p_t /
p_r$ exactly we can design networks which will be relatively robust.
For example, the bimodal network with $r=0.03$ is relatively robust for
$p_t / p_r \lesssim 0.1$ and the bimodal network with $r=0.09$ is robust
for $p_t / p_r \lesssim 1$.  Also plotted in Fig.~\ref{final2delta} is
the optimal scale-free network with $\kav=3$.  We see that the optimal
bimodal network is more robust than the optimal scale-free network and
we can even pick a configuration with fixed $r$ (e.g. $r=0.03$) which is
more robust than the optimal scale-free network in most ranges of $p_t/p_r$.

\begin{figure}[thb]
\includegraphics[width=6cm]{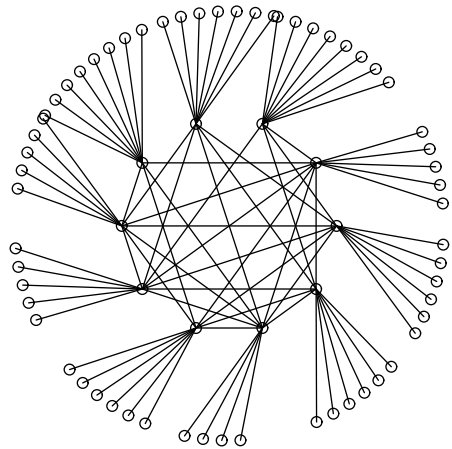}
\caption{Realization of bimodal network with $N=100$ nodes, $\kav=2.1$
   and $r=0.1$, so there are $rN=10$ ``hub'' nodes of degree 12, as found from Eq.~(\ref{eq0}).}
\label{fig:ex21}
\end{figure}

In Fig.~\ref{fig:ex21} we show a typical optimal realization of a bimodal
network. The network of $N=100$ nodes consists
of $rN$ nodes with $k=k_2$ (``hubs'') which are highly connected among
themselves; the nodes of single degree are each connected to one of
these hubs.  We note that while the hubs are highly connected among
themselves, they do not form a complete graph --- every hub is not
connected to every other hub.  For larger $N$, the fraction of hubs to
which a given hub connects decreases but the robustness of the network
is unchanged.

In summary, we have provided a qualitative argument and numerical
results which indicate that the most robust network to multiple waves of
targeted and random attacks has a bimodal degree distribution with a
fraction $r$ of the nodes having degree $k_2= (\kav - 1 +r)/r$ and the
remainder of the nodes having degree 1. The optimal value of $r$ is
approximately $1.7 \left( p_t/p_r \right)$ for $p_t/p_r\ll 1$.  For larger values of
$p_t/p_r$, the optimal value of $r$ is 1 and all nodes have degree
$\kav$.  Even if $p_t/p_r$ is not known exactly, a value of $r$ can be
chosen which maximizes the network robustness over a wide range of
values of $p_t/p_r$, as seen in Fig.~\ref{final2delta}.

We note that while the optimal distribution found here and that found in
Ref.~\cite{Paul2004} are both bimodal, the values of the parameters
characterizing these distributions are different.  As found in
Ref.~\cite{Paul2004} the network with optimal resilience to either
random or targeted attack has $r=1/N$ and $k_2 \sim r^{-2/3}$.  Finally,
we note that it is possible to prove analytically that for the case in
which a single targeted attack followed by a single random attack
results in the network becoming disconnected, the optimal distribution is
also bimodal with $k_1=1$, $k_2=(\kav -1+r)/r$ and $r$ of the order of
$p_t/p_r$~\cite{new} ---  supporting the results found here for
multiple waves of attacks.

We thank S.~Sreenivasan for helpful
discussions and ONR and the Israel Science Foundation for support.


\begin{thebibliography}{99}
\bibitem{note0} In order to keep $\langle k \rangle$ constant, as we
change $\lambda$ we also must change $m$, the minimal number of links a
node can have.  Note that the value of $m$ decreases as we increase
$\lambda$.
\bibitem{note1} We obtain similar results for other values of $\langle k \rangle$.
\bibitem{noteX} A similar dependence on only $p_t/p_r$ is also found in
other network types including scale-free.
\bibitem{noteA} A similar results are obtained for other values of $\kav$.
\end{thebibliography}
\end{document}